\documentclass[twocolumn, secnumarabic, amssymb, nobibnotes, prl, superscriptaddress,aps,prl]{revtex4-2}

\usepackage{graphicx} 
\usepackage{amsmath}
\usepackage{float}
\usepackage{color}
\usepackage[english]{babel}
\usepackage{array}
\usepackage[T1]{fontenc}
\usepackage[usenames,dvipsnames]{xcolor}
\usepackage{setspace}
\usepackage{hyperref}
\usepackage{bm}
\usepackage{lipsum} 
\usepackage{tikz}
\usepackage{stmaryrd}

\usepackage{blkarray}
\usepackage{mathtools}
\usepackage{multirow}
\makeatletter
\renewcommand*{\fnum@figure}{{\normalfont \small{FIG.}~\thefigure}}
\makeatother

\begin{document}

\title{Microscopic description of the intermittent dynamics driving logarithmic creep}

\author{Daniel Korchinski}
\altaffiliation{These authors contributed equally to this work}

\affiliation{Department of Physics and Astronomy and Stewart Blusson Quantum Matter Institute, University of British Columbia, Vancouver BC V6T 1Z1, Canada}

\author{Dor Shohat}
\altaffiliation{These authors contributed equally to this work}

\affiliation{Department of Condensed Matter, School of Physics and Astronomy, Tel Aviv University, Tel Aviv 69978, Israel}
\affiliation{Center for Physics and Chemistry of Living Systems, Tel Aviv University, Tel Aviv 69978, Israel}

\author{Yoav Lahini}
\affiliation {Department of Condensed Matter, School of Physics and Astronomy, Tel Aviv University, Tel Aviv 69978, Israel}
\affiliation{Center for Physics and Chemistry of Living Systems, Tel Aviv University, Tel Aviv 69978, Israel}

\author{Matthieu Wyart}
\affiliation{Physics Institute, École Polytechnique Fédérale de Lausanne (EPFL), CH-1015 Lausanne, Switzerland}

\begin{abstract}
Disordered materials under an imposed forcing can display creep and aging effects, accompanied by intermittent, spatially heterogeneous dynamics. We propose a unifying microscopic description of these phenomena, based on the notion that as the system ages, the density of local barriers that enable relaxation displays a slowly evolving gap. As a result, the relaxation dynamics is dominated by the activation of the lowest, extremal tail of the distribution. This framework predicts logarithmic creep, as well as correlated bursts of slow activated rearrangements, or 'thermal avalanches', whose size grows logarithmically with their duration. The time interval between events within avalanches obeys a universal power-law distribution, with a cut-off that is simply proportional to the age of the system. We show that these predictions hold both in numerical models of amorphous solids, as well as in experiments with thin crumpled sheets. This analysis suggests that the heterogeneous dynamics occurring during logarithmic creep is related to other phenomena, including dynamical heterogeneities characterising the glass transition. 
\end{abstract}
\maketitle

{\bf Introduction:} Disordered systems in even the simplest situations can give rise to rich emergent physics. Consider the behaviour of a thin sheet of plastic crumpled into a ball. This seemingly mundane system exhibits a wide range of far from equilibrium behaviours, including slow relaxations, creep and aging effects \cite{matan_crumpling_2002, lahini2017nonmonotonic}, intermittent response accompanied by crackling sounds \cite{lahini2023crackling} and avalanche statistics \cite{shohat2023logarithmic}, as well as strong memory effects \cite{matan_crumpling_2002, shohat2022memory, shohat2023dissipation, lahini2017nonmonotonic}. 
Under a change of external load, the crumpled sheet compacts logarithmically over many decades in time, exhibiting ``creep'' flow. Such logarithmic creep is ubiquitous to disordered solids, and has been observed over timescales ranging from seconds to days (or even years) including in  crumpled sheets~\cite{lahini2017nonmonotonic,lahini2023crackling,shohat2023logarithmic,matan_crumpling_2002},  disordered electronic systems \cite{woltjer_time_1993,vaknin_aging_2000,grenet_anomalous_2007}, frictional contact strength~\cite{ben-david_slip-stick_2010,rubinstein_contact_2006}, polycrystalline ice~\cite{duval_anelastic_1978},  granular compaction \cite{
nicolas_compaction_2000,murphy_memory_2020,ben-naim_slow_1998,knight_density_1995,boutreux_compaction_1997}, the motion of pinned elastic interfaces~\cite{purrello_creep_2017, kolton_creep_2005}, and in amorphous solids~\cite{popovic2022scaling,vasisht_residual_2022,liu_elastoplastic_2021,cabriolu_precursors_2019}. The experimental accessibility of crumpled sheets offers a view of the microscopic processes underlying logarithmic creep. It was recently shown that the logarithmic compaction of crumpled sheets under load advances via ever slowing, scale-free cascades of mesoscopic instabilities~\cite{shohat2023logarithmic}. Yet, A complete microscopic description of the observed intermittent and slow behavior and the resulting logarithmic aging is still missing.

Scale-free distributions of rapid activity bursts, commonly known as crackling noise \cite{sethna2001crackling}, are  commonly observed in disordered driven systems \cite{salje_crackling_2014,rosso2022avalanches}. These phenomena are mostly studied theoretically at zero temperature via mean-field approximations, scaling and renormalization group arguments \cite{Kardar1998,Fisher1998} or by considering the self-organized criticality of cellular automata \cite{bak_self-organized_1987,bak_self-organized_1988,turcotte_landslides_2004}. At finite temperatures, the situation is more complex. `Thermal avalanches' of slow activated events  have  been argued for in that case, but their nature is debated \cite{Chauve2000,de2024dynamical,tahaei2023scaling,Durin2023,ferrero2021creep,takaha_avalanche_2024}. Refs. \cite{de2024dynamical,tahaei2023scaling} predicted that thermal avalanches grow logarithmically in time and are responsible for dynamical heteorgeneities in super-cooled liquids near their glass transition ~\cite{kob1997dynamical,yamamoto1998dynamics,dalle2007spatial,karmakar2014growing,Ozawa2023,guiselin2021microscopic} and in the stationary creep flows of pinned interfaces \cite{de2024dynamical,ferrero2017spatiotemporal}. However, beyond toy models, empirical support for this view is limited to molecular dynamics of model liquids near their glass transition \cite{tahaei2023scaling,gavazzoni2023testing} or immediately after quench~\cite{takaha_avalanche_2024}, and experimental validation is missing. Moreover, this approach considered stationary conditions and its application to aging phenomena is unclear.

In this letter, we extend these views to obtain a microscopic description of logarithmic aging. Our approach, which focuses on the evolution of the distribution of barriers in the material, naturally captures the logarithmic slowdown observed in creep as well as the observed statistics of thermal avalanches. It also makes novel predictions on the distribution of time lapse between instabilities within avalanches. In a second step, we successfully test these predictions in experiments with crumpled thin sheets, and in numerical simulations of model of amorphous solids under an applied load. Overall, our work leads to a novel framework to describe aging occurring during creep, that connects microscopic and macroscopic phenomena in a variety of driven systems. 

{\bf A mesoscopic approach:} We model disordered system in a mesoscopic manner, as being composed of $N=L^d$ blocks, which satisfy the following properties. (i) Each block $i$ is characterized by its stability $x_i$, a dimensionless quantity. For example, for amorphous solids under an applied shear, $x_i$ is  often considered to represent the difference between the shear component of the local stress tensor and that of the local yield stress. (ii) A stability index $x$ implies that the block displays an energy barrier for relaxation $E=\epsilon_0 x^\alpha$. The constant $\epsilon_0$ can be taken to be unity without loss of generality for a proper choice of the scale of stability values, whereas $\alpha$ is an exponent that depends on the smoothness of the local energy barrier ($\alpha=3/2$ if the energy landscape is smooth and $\alpha=1$ if it presents cusps \cite{Middleton1992a,chattoraj_universal_2010}). At finite temperature $T$ (in units where the Boltzmann constant is $k_B=1$), each stable site can be activated and rearrange with an Arrhenius rate $\lambda(x) = \exp[-E /T]/\tau_0$. $\tau_0$ is a microscopic time scale that henceforth defines our units of time, such that $\tau_0=1$. (iii) Sites are elastically coupled with some interaction kernel $G(\vec{r}_{ij})$ describing the spatial redistribution of stress after a relaxation event. $G(\vec{r}_{ij})$ can be local and only couple nearby sites, or can be long-ranged. For amorphous solids, this kernel generically decays as $G(\vec{r}_{ij})\sim 1/r^d$ and can change sign depending on the direction considered \cite{Picard2004}. Such a mesoscopic view has a long history in various fields including the depinning transition \cite{Middleton1992a,Fisher1998,ferrero2021creep}, the glass transition \cite{Argon1995,guiselin2021microscopic,Heuer2008,Ozawa2023,tahaei2023scaling},  the plasticity of amorphous solids \cite{Picard2004,ferrero2014relaxation,popovic2021thermally} or crumpled thin sheets \cite{shohat2022memory,shohat2023logarithmic}.

{\bf Existing arguments for stationary flows:} At vanishing temperature $T=0^+$,  the dynamics in this viewpoint is approximately `extremal': the weakest site (i.e. the smallest $x_i$ value, which we denote $x_{min}$) relaxes first. Extremal dynamics has been studied extensively in the context of self-organized criticality \cite{bak_self-organized_1987,bak_self-organized_1988,paczuski1996avalanche}. It has been shown that for stationary conditions, as $N\rightarrow  \infty$, the distribution of stability $P(x)=\sum_i \delta(x-x_i)/N$ develops a gap at a finite stability value $x_g$. Yet, below the gap there is a sub-extensive population of stable sites. In the thermodynamic limit, the smallest stability value $x_{min}$ is distributed as:
\begin{equation}
\label{e1}
    P(x_{min})\sim (x_g-x_{min})^{-\beta}
\end{equation}
where $\beta$ is some exponent \cite{popovic2021thermally}. As is the case for all exponents introduced in this work, its value can depend on the shape of the propagator 
$G(\vec{r}_{ij})$. 

\begin{figure}
    \centering 
    \includegraphics[width=0.48\textwidth]{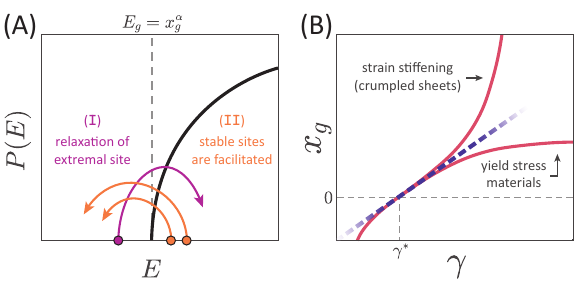}
    \caption{\textbf{Theoretical scenario} - (A) Dynamical facilitation in the density of local barriers, which characterizes the energy landscape. The extremal site with lowest energy (typically close to $E_g$) is activated by thermal fluctuations. It facilitates stable sites to lower energies. These sites are then activated quickly, altogether forming a thermal avalanche; (B) Sketch for the stress gap $x_g$ as a function of strain $\gamma$ at  small temperatures and fixed external stress $\sigma$. 
    For $T=0^+$, extremal dynamics is a good approximation, and leads to a stability gap $x_g$ that is zero at $\gamma = \gamma^*$ and initially grows linearly with strain.
    The linear approximation of $x_g(\gamma)$ close to $\gamma^*$ (dashed line), which lies at the base of our theory, can be applied both to strain stiffening materials and to yield stress materials.}
    \label{fig:theory}
\end{figure}

At finite temperature, it was argued in \cite{de2024dynamical} that extremal dynamics remains a good approximation below some length scale $\xi\sim T^{-\nu}$ characterizing dynamical heterogeneities. For smaller systems $L<\xi$ as considered in what follows, thermal avalanches can be defined as cascades of thermally activated events driven by facilitation \cite{tahaei2023scaling}. In essence, each local instability can lower the barriers of neighboring stable sites, facilitating their thermal activation \cite{paczuski1996avalanche}. This is illustrated in Fig. \ref{fig:theory}A.
For any arbitrary threshold time $t_0$, sequences of $S$ events separated by time interval shorter than $t_0$ are power-law distributed with:
\begin{equation}
\label{eq:scalingava}
P(S)\sim S^{-\tau}g(S/S_c)
\end{equation}
Where $g(z)$ is a cutoff function that decays sharply from $1$ to $0$ around $z\approx1$. The cut-off $S_c$ satisfies:
\begin{equation}
   S_c\sim [T\ln(t_{\text{sta}}/t_0)]^{-\nu d_f}\,,\label{eq:av_cutoff_glass}
\end{equation}
where $d_f$ is the fractal dimension. Here $t_{\text{sta}}\equiv \exp[E_g/T]$ is the characteristic relaxation time of the system in the stationary state. In essence, thermal avalanches are anomalously slow, and their size grows logarithmically with the cut-off time interval $t_0$ that defines them. Because the overall duration of an avalanche is dominated by the longest intervals between events constituting it, this duration is also of order $t_0$. Eq.\ref{eq:av_cutoff_glass} was never tested empirically, a test we perform below using experiments on crumpled sheets. 

{\bf Extension to aging dynamics:} 
Consider a system where a load is imposed at $t=0$. In the limit of small temperatures, the dynamics will be approximately extremal, as least stable sites will yield much faster than others. We thus expect a gap $x_g$ to rapidly form; yet its value must slowly evolve to adapt to the applied load. This view is consistent with experiment and modeling of crumpled sheets, for which the system's relaxation slows down due to a slow but steady increase in the system's lowest (local) energy barrier \cite{shohat2023logarithmic}. %
Thus, the dynamics will be similar to the stationary case described above: microscopic relaxation is characterized by long waiting times of order $\exp(x_g^\alpha/T)$ at which thermal avalanches are triggered. Here, however, $x_g$ evolves in time, thus accounting for the aging-related slowdown.

{\it Logarithmic creep:} To characterize this evolution, we denote by $\gamma$ the quantity conjugate to the applied load (corresponding to the strain in amorphous solids or crumpled sheets). $x_g(\gamma)$ must be a smooth function in the extremal dynamics approximation. Denoting $\gamma^*$ the strain at which mechanical stability is reached, i.e. $x_g(\gamma^*)=0$, we must then have $x_g(\gamma)\approx C_0 (\gamma-\gamma^*)$ in the neighborhood of $\gamma^*$. This is illustrated in Fig. \ref{fig:theory}B.
The strain rate $\dot\gamma$ is inversely proportional to the long waiting time between thermal avalanches set by $x_g$, and thus must follow at the leading order:
\begin{equation}
\label{e3}
 \dot{\gamma} \sim  \exp(-(\gamma-\gamma^*)^\alpha C_0^\alpha/T) 
\end{equation}
In the Appendix~\ref{app:gap_evolution} we show that this equation leads approximately to 
\begin{eqnarray}
\label{e4}
\gamma-\gamma^* &\sim& [T\log(t)]^{1/\alpha} \\ \dot{\gamma}&\sim& \frac{1}{t} 
\end{eqnarray}
where logarithmic pre-factors are neglected in the last expression. This analysis thus recovers the logarithmic response observed as such systems age. From Eq.\ref{e4} we also obtain for the gap energy controlling the relaxation time:
\begin{eqnarray}
\label{e5}
E_g &\sim& T\log(t)
\end{eqnarray}

This derivation is reminiscent of phenomenological strain hardening models used to account for logarithmic creep \cite{matan_crumpling_2002,jaeger1989relaxation}, by assuming that the system's evolution is dictated by a single energy barrier that grows in proportion to the overall strain. Here, this is derived from first principles as a result of extremal activation of the distribution of barriers near the gap.

{\it Thermal avalanches:} The increasing stability gap leads to a growing timescale $t_g\equiv \exp[-E_g(\gamma)/T]$.  Avalanches being sub-extensive, the effect of a single avalanche on the gap (an intensive quantity) must be vanishing in the thermodynamic limit. Thus as far as a single avalanche is concerned, the gap is essentially constant in time.
The results on avalanche statistics in the stationary case must then apply: these avalanches will also exhibit a cut-off $S_c$ as in eq.~\ref{eq:av_cutoff_glass}, with the evolving $t_g$ playing the role of $t_{\text{sta}}$, so that 
\begin{equation}
\langle S_c\rangle\sim [T\ln( t_g/t_0)]^{-\nu d_f}\,,
\label{eq:ava_dist_cutoff_nolog}
\end{equation}
with pre-factors independent of $t_0$ that slowly grow in time before saturating, as detailed in the appendix.

\begin{figure*}
    \centering
    \includegraphics[width=1\textwidth]{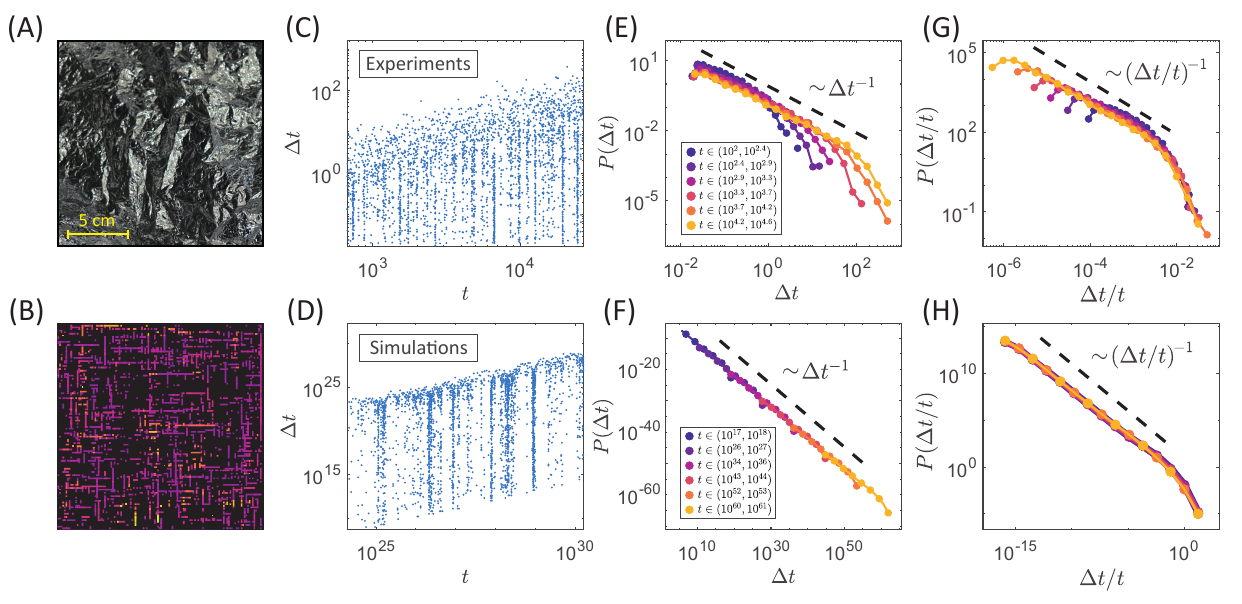}
    \caption{\textbf{Temporal statistics of instabilities} - Instabilities during logarithmic creep in experiments (top) and simulations with $L=128$ (bottom): (A) An unfolded crumpled sheet of mylar; (B) Visualization of the EPM's activity over a strain interval of $\Delta \gamma \approx 0.4$. Instability locations are colored by the log of their occurrence time, revealing many line-like avalanches; (C,D) Waiting times $\Delta t$ between instabilities as a function of time. Each point represents a single instability; (E,F) Waiting time distributions $P(\Delta t)$ in different time windows going from purple to yellow. All distributions follow $P(\Delta t)\sim\Delta t^{-1}$; (G,H) Normalizing by $t$, the distributions $P(\Delta t/t)$ collapse to a single curve for the same time windows. }
    \label{fig:waiting_times}
\end{figure*}

{\it Waiting time distribution:}
During thermal creep instabilities follow universal temporal statistics. Assuming extremal dynamics, so that creep is dominated by the waiting time of the weakest site, with $P(E_{min})\sim (E_g-E_{min})^{-\beta}$ from eq.~\ref{e1}, then $P(\Delta t/t_g) = \int P(\Delta t/t_g|U) P(U) dU $ where $U=E_g-E_{min}$. Since the waiting time $\Delta t$ distribution for a site with barrier $E_{min}$ is exponential, we have that $P(\Delta t/t_g|U)= \lambda e^{-\lambda \Delta t/t_g}$    %
for Arrhenius rate $\lambda = \exp(U / T)$. For $\beta=0$ an exact integration is possible, leading to:
\begin{equation}
\label{eq:waiting_times}
 P(\Delta t/t_g)\approx \left(\frac{\Delta t}{t_g} \right)^{-1}\exp\left[-\frac{\Delta t}{t_g}\right] \,,
\end{equation}
which holds for $\Delta t\gg1$, and where a logarithmic prefactor in $t_g$ is neglected, as reported in the Appendix. This formulation demonstrates that the distribution of $\Delta t/t_g$ essentially does not depend on $t_g$, as we will confirm empirically below.
When $\beta>0$, there are additional logarithmic corrections (see appendix for a derivation).

{\bf Experiments and simulations:}
To verify our theory, we consider the experimental system described in Ref. \cite{shohat2023logarithmic}. Thin sheets of Mylar, $50\times50\,$cm across and $8\,\mu$m thick, are crumpled manually and placed under an external load $M=400\,$g while their height $h$ is monitored. Over time, a steady logarithmic compaction $h=h_0-b\ln(t)$ is observed. As the sheet compacts it emits crackling noise \cite{kramer1996universal,lahini2023crackling}, intermittent acoustic pulses which are picked up by a nearby microphone and assigned a timestamp $t_i$. These originate from mesoscopic snap-through instabilities in the sheet \cite{shohat2022memory,shohat2023dissipation}, which were shown to govern the logarithmic compaction dynamics \cite{shohat2023logarithmic}.

We complement the experiments with a minimal mesoscopic thermal elastoplastic model (EPM) of amorphous plasticity, as described in~\cite{korchinski_dynamic_2022,korchinski_thermally_2024}. The model consists of a grid of $L^2$ elastically coupled sites. Each site is associated with a local stress threshold $\sigma_{th,i}$ (sampled from a Weibull distribution with shape parameter $k=2$) that, when exceeded, allows for plastic rearrangement in the site. These localized rearrangements allow the site to locally strain (by $\approx L^{-d}\delta \gamma$) into a more favourable configuration, reducing the local stress. This induces a long-range stress redistribution throughout the system, stabilizing some sites and destabilizing others, with an Eshelby-like kernel $G(\vec{r}_{ij})\sim \cos(4\theta)/r^d$ computed using finite elements, where $\theta$ is the angle between $\vec{r}_{ij}$ and the direction of applied external shear. We allow for the thermal activation of otherwise mechanically stable sites, (i.e. with $x_i = \sigma_{th,i} - \sigma > 0$) with an Arrhenius rate $\lambda(x) = \exp[-x^\alpha /T]$, we choose $\alpha = 3/2$, as predicted from catastrophe theory for smooth particles interactions \cite{chattoraj_universal_2010,maloney_energy_2006}, and $T$ a rescaled temperature of order $10^{-3}$. We load our EPM simulations at a fixed stress $\Sigma = 0.5$, well below the critical flow-stress $\Sigma_c \approx 0.73$ for our model. Avalanches and rheology of this model have been characterized near the critical flow-stress at finite-temperature in~\cite{korchinski_thermally_2024}. 

Despite the distinct differences in the rearrangement mechanisms, the interactions, and the fate under large deformation (namely strain stiffening vs flowing), both experiments and simulations are captured by our theoretical analysis. They exhibit logarithmic thermal creep via avalanches of instabilities, corresponding to an opening gap in the distribution $P(x)$. Most prominently, this dynamics is evident when plotting the waiting time between instabilities $\Delta t_i=t_i-t_{i-1}$ vs. their occurrence time $t_i$ \cite{shohat2023logarithmic}, as shown in Fig. \ref{fig:waiting_times}C,D. During creep the typical waiting time between instabilities grows, yet we observe thermal avalanches of instabilities which involve long timescales. These are indicated by the vertical streaks in Fig. \ref{fig:waiting_times}C,D. Fig. \ref{fig:waiting_times}B shows an example of the spatiotemporal dynamics of activity  in the EPM, with characteristic ``line-like'' avalanche imply a fractal dimension $d_f\approx 1$. From the envelope of the stabilities and waiting times between instabilities, we can access the gap $x_g$ and $t_g$ (see appendix for detailed protocol).

{\bf Agreement with theory:}
We begin by considering waiting time distributions $P(\Delta t)$. We divide the experiments and simulations to logarithmically spaced intervals, such that in each bin the cutoff $t_g$ is approximately constant. As predicted by extremal dynamics of Arrhenius crossings, $P(\Delta t)$ in each interval follows the universal power-law of Eq. \ref{eq:waiting_times}, as shown in Fig. \ref{fig:waiting_times}E,F. Indeed, the logarithmic corrections to Eq. \ref{eq:waiting_times} can be neglected. The cutoff grows linearly with time $t_g\sim t$, and the distributions over different intervals collapse when normalizing by time and calculating $P(\Delta t/t)$, as shown in Fig. \ref{fig:waiting_times}G,H.

\begin{figure}
    \centering 
    \includegraphics[width=0.48\textwidth]{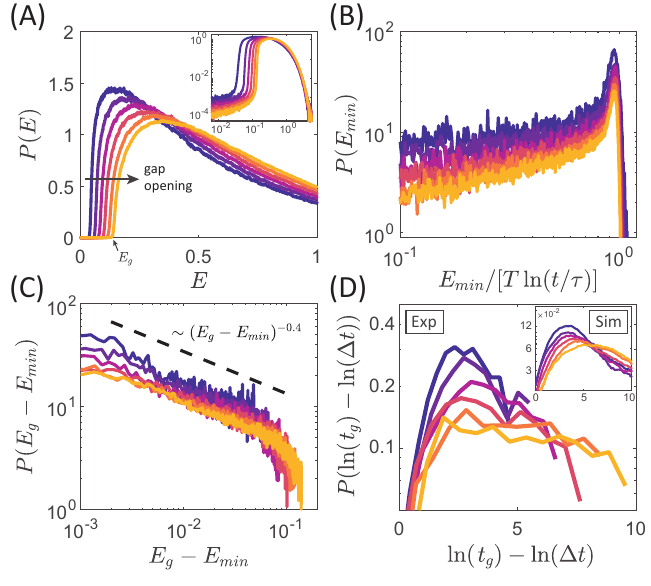}
    \caption{\textbf{Evolution of the energy landscape} - (A) Full energy barrier distributions $P(E)$ across different ages in simulations. With time, the distribution develops a gap $E_g$. Color coding is consistent with the time windows in Fig. \ref{fig:waiting_times}. The inset shows the same plot in log-log scale; (B) Barrier distributions for failing sites only $P(E_{fail})$, as a function of $E_{fail}$ normalized by the gap $E_g\sim T\ln(t)$. The collapsed curves reveal rich activity below the gap $E_g$ the curves at different times; (C) Barrier distribution of failing sites relative to the gap, exhibiting a power-law $P(E)\sim (E_g-E)^{-\beta}$ with $\beta\approx0.4$; (D) Barriers can be estimated from waiting times as $E/T = \ln(\Delta t)$, leading to similar results in the experiments and the simulations (inset).%
    }
        \label{fig:landscape}
\end{figure}

Next, we consider the energy landscape and its evolution during creep. The simulations allow accessing the full barrier distribution $P(E)$ at a given time. As predicted, for low $E$ the distribution develops a gap over time, as shown in Fig. \ref{fig:landscape}A. Note that the usual power-law behavior $P(E)\sim (E-E_g)^{\tilde{\theta}}$ is observed above the gap, as expected in the presence of long-range interactions with varying sign \cite{lin2016mean,muller2015marginal,lin2014scaling}. Importantly, below this gap the distribution is nonzero, though fluctuating and sub-extensive with system size (inset). Creep is dominated by these barriers below the gap $E_g$. Indeed, isolating only the density of instabilities that were triggered $E_{fail}$ reveals a wide distribution peaking at $E_g(t)$, as shown in Fig. \ref{fig:landscape}B. As detailed in Appendix, we use statistical methods to extract the gap magnitude $E_g$ and confirm Eq.\ref{e5}: $E_g\sim T\ln(t)$. We find that the barriers are power-law distributed relative to the gap $P(E)\sim (E_g-E)^{-\beta}$ with $\beta\approx0.4$, as shown in Fig. \ref{fig:landscape}C. This result is consistent with recent simulations of thermally activated, disordered interfaces, where a power-law distribution of extremal energy barriers develops below a fixed energy gap \cite{de2024dynamical}.

In experiments, the distribution of failing sites can only be accessed indirectly, by approximating $E_{fail} \sim\ln(\Delta t)$ and $E_g \sim \ln(t_g)\sim\ln(t)$. In contrast to the actual barrier energy distribution, the distribution relative to the gap $P(\ln(t_g)-\ln(\Delta t))$ exhibits a peak and a sharp decay close to zero, as shown in Fig. \ref{fig:landscape}D. This effect originates from fluctuations in the barrier crossing time $\Delta t$: although a barrier of height $E$ predicts a mean waiting time of $\tau \exp(E/T)$, the waiting time distribution is exponential, meaning that longer times are possible. 
Since fluctuations are in the order of the mean crossing time, $\ln(t_g)$ is in fact an over-estimation of the gap when compared to extracting it directly. Indeed, the same distribution $P(\ln(t_g)-\ln(\Delta t))$ exhibits a peak in the simulations as well, and resembles the experimental results (Fig. \ref{fig:landscape}D inset). This effect hinders the ability to extract the logarithmic correction $\beta$ from experimental data, but the direct correspondence between numerics and experiments support our approach.

\begin{figure}
    \centering
    \includegraphics[width=0.48\textwidth]{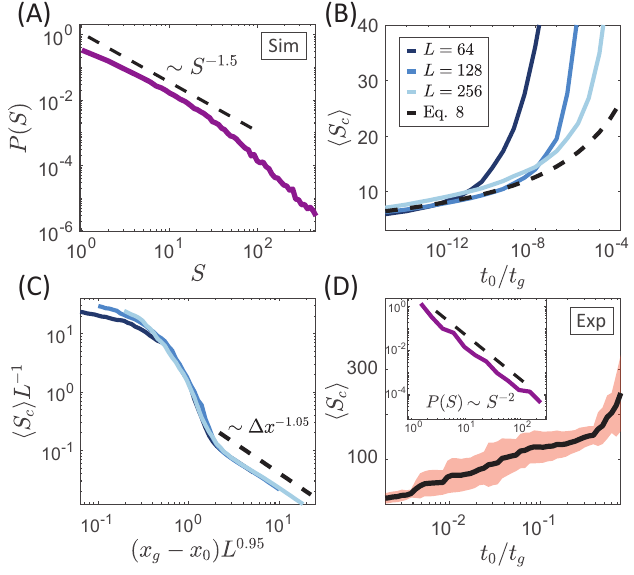}
    \caption{\textbf{Thermal avalanches} - Each avalanche is calculated as the sequences of events with $\Delta t_i<t_0<t_g$, while the gap is $t_g=Ct$. (A) Avalanche distribution in simulations with $L=256$, and for fixed $t_0/t_g=10^{-4}$, exhibiting a power-law decay $P(S)\sim S^{-\tau}$, with $\tau=1.5\pm 0.1$. The exponent $\tau$ is obtained via maximum likelihood estimation; (B) Avalanche cutoff $\langle S_c\rangle$ as a function of $t_0/t_g$ in simulations, and a fit to Eq. \ref{eq:ava_dist_cutoff_nolog} with $\nu d_f\approx 1.05$. As system size is increased, the fit holds closer to the gap and for larger $t_0/t_g$; (C) Finite-size scaling of the avalanche cutoff, plotted by the distance to the stability gap $\Delta x=x_g-x_0$. This reveals a minimum stability scale $\sim L^{-0.95}$, above which our theory holds; (D) Thermal avalanches in experiments. The inset shows the avalanche distribution for $t_0/t_g=10^{-1}$. The distribution exhibits a power-law with $\tau=2\pm0.2$. The main panel shows the avalanche cutoff $\langle S_c\rangle$ as a function of $t_0/t_g$ (solid curve), which agrees qualitatively with the theoretical prediction. The shaded region marks the fluctuations between experiments.}
    \label{fig:avalanches}
\end{figure}

We now examine the distribution of thermal avalanches. The magnitude of each avalanche $S$ is calculated as the number of consecutive events with $\Delta t_i<t_0$. Whenever $t_0$ is exceeded, a new avalanche begins. Crucially, $t_0$ is not constant but instead grows linearly with time $t_0=At$, such that the distance to the gap $t_0/t_g$ remains constant during aging. Indeed, for $t_0/t_g<1$ we find that the avalanche distributions are well captured by eq.\ref{eq:scalingava}, and exhibit scale free distributions with $\tau\approx2\pm 0.2$ in experiments and $\tau\approx1.5 \pm 0.1$ in simulations (as determined by maximum-likelihood estimation~\cite{bauke_parameter_2007}). This is shown in Fig. \ref{fig:avalanches}A,D.

We can now test the prediction of Eq.~\ref{eq:ava_dist_cutoff_nolog}. We vary $t_0/t_g$ systematically, and compute the thermal avalanche distribution for each value. The distribution cutoff $\langle S_c\rangle$ is obtained from the moment ratio $\langle S_c\rangle\sim\langle S^{k+1}\rangle/\langle S^k\rangle$\footnote{For simulations, we use $k=1$ to reduce error, but for experiments we use the higher moment $k=2$ because the $p(s)$ distribution has an exponent $\tau > 2$ for small $t_0/t_g$.}. As a function of $t_0/t_g$, it exhibits the predicted slow logarithmic growth until diverging close to $t_0/t_g\rightarrow1$. This is well captured by Eq.~\ref{eq:ava_dist_cutoff_nolog} in the numerical model, as shown in Fig. \ref{fig:avalanches}B. Here, we find $\nu d_f\approx 1.05$ consistently with previous studies \cite{lin2014scaling}.

This fit succeeds only up to a system-size dependent ratio $t_0/t_g < r_{\rm coal.}(L)$, which marks the onset of coalescence between different thermal avalanches. In other words, the ability to separate avalanches is limited by finite-size fluctuations, and as $t_0 \rightarrow t_g$, these fluctuations lead to the coalescence of independent cascades. Fig.~\ref{fig:avalanches}C shows how larger systems can probe the dynamics closer to the gap, extending the  $s_c \sim \Delta x^{-1/\sigma}$ scaling to a minimum stability scale $x_g - x_0 \sim L^{-0.95}$.  Accordingly, very large avalanches are formed exclusively through coalescence and belong to a distinct percolation universality class that is an artifact of our measurement strategy in finite-systems. A similar transition occurs in neuronal avalanches when coalescence becomes dominant~\cite{korchinski_criticality_2021}. In the appendix, we address this with a full finite-size scaling analysis.

For the experiments, the growth of the cutoff $\langle S_c\rangle$ qualitatively matches the expected behavior. It exhibits a slow growth and diverges when $t_0/t_g\rightarrow1$. However, finite-size fluctuations along with our temporal resolution do not allow a precise estimation of $\nu d_f$ or a fit to Eq. \ref{eq:ava_dist_cutoff_nolog}. %

{\bf Discussion:}
We have proposed a microscopic framework for the aging and creep flows of various disordered materials. In this description, a key quantity is the distribution of local barriers, and in particular its extermal tail. We have described how this density adiabatically ages in time by opening up a gap. This view naturally explains ubiquitous observations, including the macroscopic logarithmic response to an imposed perturbation, and the presence of an intermittent microscopic dynamics. Concerning the latter, 
we predicted  that it occurs by burst of slow activated events, whose stability is smaller than that of the gap. Quantitatively, the size of these bursts is logarithmic in their duration. The distribution of time laps between events collapses when rescaled by the system age, and follows an inverse power-law. 
The vast differences between the systems we study, crumpled this sheets and amorphous solids, suggest the universality of our description. 

Central to our description is the appearance of a time-dependent gap in the distribution of energy barriers, and the recognition that creep proceeds through the activation of sites below this gap. Thus, the distribution of activated barriers has a sharp cutoff $P(E_{min}>E_g)=0$. The weakest energy barrier is essentially never above $E_g$ owing to the rep-population of sites below the gap by facilitation.
This contrasts with renewal models such as Bouchaud's trap model~\cite{bouchaud_weak_1992}, where the exponential tail of a \textit{fixed} barrier distribution $\exp(-E_{min}/T_g)$ gives rise to a power-law distribution of waiting times, and aging. Alternative descriptions of logarithmic creep consider a wide range of independent relaxation modes \cite{amir2012relaxations}, leading to log-Poisson statistics \cite{lahini2023crackling}. While the exponential cutoff of Eq. \ref{eq:waiting_times} is reminiscent of log-Poisson dynamics \cite{boettcher2023instability}, the approximately inverse power-law reveals that the dynamics is far more complex and highly correlated.

Our scenario incorporates the notion that dynamical heterogeneities are induced by logarithmically-growing thermal avalanches. This view has received numerical support in stationary conditions in systems as diverse as pinned interfaces \cite{de2024dynamical} or  molecular liquids near their glass transition \cite{gavazzoni2023testing}. Our work gives the first empirical test for the dynamics of thermal avalanches, in addition to extending this approach to non-stationary conditions. Such heterogeneous dynamics may thus be generic to a broad class of disordered materials and phenomena.

Our framework offers new perspectives to study complex aging phenomena, including non-monotonic aging \cite{lahini2017nonmonotonic,kovacs1963glass,riechers2022predicting}, memory \cite{josserand2000memory,lahini2017nonmonotonic,dillavou2018nonmonotonic}, and rejuvenation effects \cite{scalliet2019rejuvenation}. 
Which observable in these materials retains the history of past perturbations imposed on them? A natural hypothesis that stems from our approach is that memory is encoded in the extremal tail of the distribution of local barriers. Finally, a central question in both material science and geophysics is what events can nucleate rupture. Sudden rupture events can occur for example when very stable amorphous materials are prepared and then loaded \cite{Ozawa18_1, popovic2018elastoplastic}, or when velocity instabilities are induced by velocity-weakening effects as in faults or frictional interfaces \cite{rubino2022intermittent}. An interesting hypothesis to investigate in the future is the notion that thermal avalanches may act as seeds that nucleate shear bands or crack-like propagating ruptures. 

{\bf Acknowledgements:}
We thank Marko Popovic for insightful discussions. We also thank Jack Parley, Alberto Rosso, and Misaki Ozawa for their helpful comments on the manuscript.
This work was supported by the Israel Science Foundation grant 2117/22.
D.S. acknowledges support from the Clore Israel Foundation. D.K. acknowledges support from the National Science and Engineering Research Council of Canada.

\bibliographystyle{ieeetr}
\bibliography{bibli}

\appendix
\section{Cut-off of avalanche sizes}
 In simulations, we  consider avalanches defined as sequences of events for which the minimal stability $x_{min}\leq x_0$.  The cut-off for  the avalanche distribution is known to scale as \cite{paczuski1996avalanche}:
\begin{equation}
    s_c\sim (x_g-x_0)^{-\nu d_f}\,. \label{app:eq:sc_xc_x0}
\end{equation}
For small enough temperatures in a finite system, an equivalent formulation allows us to define avalanches strictly in terms of experimentally accessible waiting-times: a $t_0$ avalanches is activity proceeding without any quiescent periods longer than $t_0$. %
We have:

\begin{equation}
    \log(t_g/t_0) = \exp[(x_g^\alpha-x_0^\alpha)/T].
\end{equation}
Using that at first order $x_g^{\alpha} - x_0^\alpha \approx \alpha x_g^{\alpha-1}(x_g-x_0)$ leads to:
\begin{equation}
    \log(t_g/t_0) = \exp[(x_g^\alpha-x_0^\alpha)/T]\approx \exp[\alpha x_g^{\alpha-1}(x_g-x_0)/T]
\end{equation}
Rearranging, and using eq.~\ref{app:eq:sc_xc_x0} to eliminate $x_g-x_0$, we obtain 
\begin{equation}
    s_c\sim [T\log(t_g/t_0)]^{-\nu d_f}x_g^{\nu d_f(\alpha-1)}
\end{equation}
In a stationary state or when approaching it, $x_g$ is approximately constant.
At early time during creep we have instead $x_g \approx [T\log(t_g)]^{1/\alpha}$, indicating that avalanches may tend to grow (logarithmically) in time if $\alpha>1$.

\section{Gap evolution solution\label{app:gap_evolution}}

\begin{figure}
    \centering
    \includegraphics{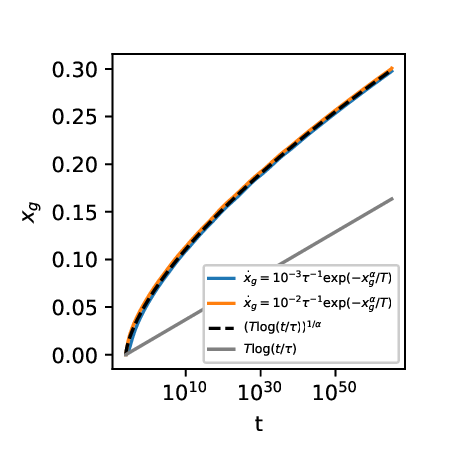}
    \caption{Numerical integration of equation~\ref{e3} yields a solution $x_g \approx [T \log(t)]^{1/\alpha}$. }
    \label{fig:numerical_xc}
\end{figure}

The equation for the growth of strain justified in the main text follows  Eq.\ref{e3}.
We are not aware of a closed form solution of that equation when $\alpha\neq 1$. It is easy to check that the  form of Eq.\ref{e4} approximately solves it, except for subdominant logarithmic terms.
The numerical solution of   Eq.\ref{e3} is shown in fig.~\ref{fig:numerical_xc}, confirming the validity of this approximation.

\section{Waiting time distribution}

Recall that the time interval distribution, normalized by $t_g$, follows $P(\Delta t/t_g) = \int \lambda e^{-\lambda \Delta t/t_g} P(U) dU $ where $U=E_g-E_{min}$, $P(U) = \frac{1-\beta}{E_g^{1-\beta}} U^{-\beta}$ and $\lambda = \exp(U / T)$.  Thus  $P(\Delta t/t_g) = \int  e^{U/T-\exp(U / T) \Delta t/t_g} U^{-\beta} dU $.    %
For $\beta=0$ an exact integration is possible, yielding
\begin{equation}
    p\left(\frac{\Delta t}{t_g}\right) = \frac{T}{E_g} \left(\frac{\Delta t}{t_g}\right)^{-1}\left[ \exp\left(-\left(\frac{\Delta t}{t_g}\right)\right)  - \exp\left(-\Delta t\right)  \right]
\end{equation}
The second exponential is only relevant when $\Delta t \lessapprox 1$, but is necessary to ensure that the distribution is normalizable. The overall normalization is maintained by the factor of $T/E_g$, and noting that $T/E_g  = \log(t_g)$, this gives the simpler expression
\begin{equation}
    p\left(\frac{\Delta t}{t_g}\right) \approx \frac{1}{\log(t_g)} \left(\frac{\Delta t}{t_g}\right)^{-1} \exp\left(-\left(\frac{\Delta t}{t_g}\right)\right)\,.
\end{equation}
For our experimental and numerical data, we cannot measure $\Delta t/t_g$ below some $\bar{t}_\textrm{min} > 1$. This means that the overall normalization does not depend on $1/\log(t_g)$, but instead on $\log(1/\bar{t}_\textrm{min})$. 

For the case $\beta \in [0,1)$ we resort to a Laplace approximation, where the integral is estimated at the value where the exponential contribution is maximum. To lowest order, this occurs at $U^*= T\log(t_g/\Delta t)$. In the limit of $\Delta t/t_g\ll 1$, we find approximately:
\begin{equation}
    p(\Delta t/t_g) \sim \frac{\left(\frac{\Delta t}{t_g}\right)^{-1} \left(T\ln\left(\frac{t_g}{\Delta t}\right)\right)^{1-\beta}}{\sqrt{ \left(\ln\left(\frac{t_g}{\Delta t}\right)\right)^{2} - \beta}}
    \label{eq:dt_no_correction}
\end{equation}
This expression is valid in the limit that $T\beta \ll 1$, $\Delta t / t_g \ll 1$, and $\Delta t \gg 1$. 

\section{Gap estimation in experiments and simulations}

To precisely assess the stability gap $x_g$ in simulations, we consider the time series of failing sites $(t_i,x_i)$ and calculate its envelope. Namely, at time $t$ the gap is 

\begin{equation}
    x_g(t)=\max_{t_i\leq t}\{x_i\}\label{eq:x_g_envelope}
\end{equation}

This property is averaged over 60 realizations, resulting in a smooth, linearly growing stability gap. The gap energy in Fig. \ref{fig:landscape} is simply $E_g=x_g^{\alpha}$.

Similarly, for the waiting times in Fig. \ref{fig:landscape}d and Fig. \ref{fig:avalanches}, we compute the envelope

\begin{equation}
    t_g(t)=\max_{t_i\leq t}\{\Delta t_i\}
    \label{eq:dt_gap_method}
\end{equation}

In experiments there is a larger variability between different realizations. Namely, the prefactor $C$ for the linear gap growth $t_g=Ct$ is not constant across all experiments. Thus instead we calculate the gap by Eq. \ref{eq:dt_gap_method} for each realization separately, and perform a linear fit to obtain a smoothly growing gap.

An alternative method to assess the growth of the gap is by the waiting time distribution cutoff. This can be accessed from the moment ratio $t_c\sim\langle \Delta t^3\rangle/\langle \Delta t^2\rangle$. Calculating the moment ratio at different ages also yields a linearly growing gap. Yet, the prefactor for the moment ratio depends on the waiting time PDF, including the logarithmic correction $\beta$. Due to this sensitivity we choose the first method which is robust regardless of $P(\Delta t)$. 

\section{Finite size scaling of the gap}
\begin{figure*}
    \centering
    \includegraphics[width=0.75\linewidth]{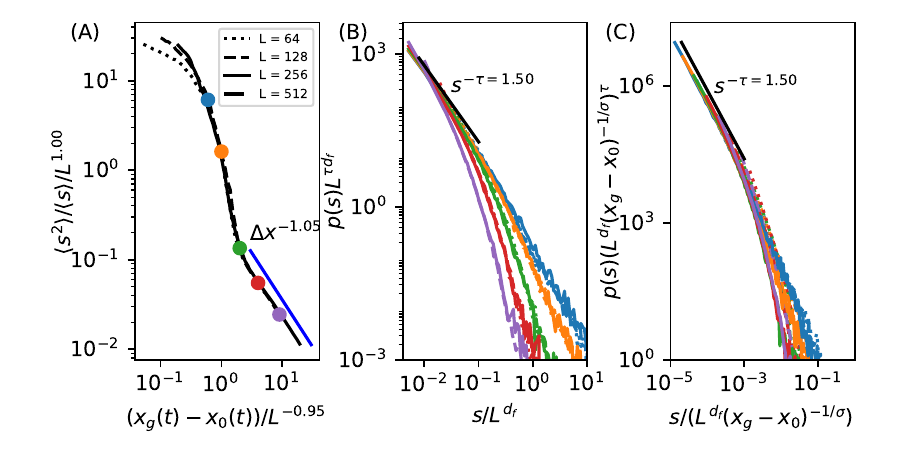}
    \caption{(A) The growth of $s_c(x_0)$ as a function of $\delta x_g$, rescaled to effect a finite-size scaling collapse. (B) Avalanche distributions, evaluated for different system sizes and at different values of $\delta x_g / L^{-a}$, also rescaled to effect a collapse. (C) The same avalanche distributions, rescaled by $s_c \sim \delta x_g^{-1/\sigma}$ to effect a collapse.}
    \label{fig:sc_x0_fss}
\end{figure*}

When the stability gap is measured using the envelope (eq.~\ref{eq:x_g_envelope}) in a finite-system, there will be fluctuations of order $\Delta x_g(L) \sim L^{-a}$ that make it impossible to accurately delineate thermal avalanches below this scale. 
That is, in an infinite system, $x_0$ avalanches exhibit a cutoff scaling as 
$s_c \sim (x_g -x_0)^{-1/\sigma}  = \delta x_g^{-1/\sigma}$. 
In a finite system, this scaling holds only for $\delta x_g > \Delta x_g(L)\sim L^{-0.95}$, as can be seen in Fig.~\ref{fig:sc_x0_fss}A. For $\delta x_g < \Delta x_g(L)$, the $x_0$ avalanches begin to merge together into much larger events. These events don't represent causal cascading activity from a single activation of an initiating site near $x_0$, but rather the merging of several consecutive $x_0$ avalanches near the fluctuating $x_g(t,L)$ gap. The rapid growth of $s_c$ for $x_g - x_0 < \Delta x_g(L)$ reflects a transition to this new merging mode of growth. Accordingly, the avalanche distribution (Fig.~\ref{fig:sc_x0_fss}B) exhibits a two powerlaw decay. The first powerlaw relates to the  causal avalanches of sites below the gap, while the second captures the effect of randomly merging together cascades. Similar distributions are encountered in neural systems when independently initiated power-law cascades (directed percolation) merge together with a distinct universality class (undirected percolation)~\cite{korchinski_criticality_2021}. The scaling of $s_c$ with $\Delta x_g^{-1/\sigma}$  only applies to the initial cascades, which is why the collapse in Fig.~\ref{fig:sc_x0_fss}C only applies to systems with $\Delta x_g > 0$. 

\end{document}